\documentclass[superscriptaddress,aps,prd,reprint,twocolumn,amsmath,amssymb,floatfix]{revtex4-2}
\usepackage{multirow}
\usepackage{lineno}
\usepackage{graphicx}
\usepackage{bm}
\usepackage{booktabs}
\usepackage{subfigure}
\usepackage{verbatim}
\usepackage[colorlinks,
            linkcolor=red,
            anchorcolor=blue,
            citecolor=cyan
            ]{hyperref}
\usepackage{orcidlink}
\usepackage[T1]{fontenc}

\begin{document}

\begin{abstract}
The precise measurement of the muon anomalous magnetic moment $a_\mu$ provides a sensitive probe of exotic interactions between muons mediated by light beyond the Standard Model (BSM) bosons.
Recent advances in both experiment and theory have largely reconciled the long-standing discrepancy in $a_\mu$.
Using the latest result, $\Delta a_\mu = a^{\rm exp}_\mu - a^{\rm SM}_\mu = (38 \pm 63) \times 10^{-11}$, we derive updated limits on light BSM boson-mediated exotic muon interactions.
\end{abstract}

\title{New Limits on Exotic Muon Interactions Mediated by Axion-Like Particles}
\author{L. Y. Wu \orcidlink{0009-0003-7568-008X}}
\email[]{22110200026@m.fudan.edu.cn}
\affiliation{Institute of Fundamental Physics and Quantum Technology, and School of Physical Science and Technology, Ningbo University, Ningbo, Zhejiang 315211, China}
\affiliation{Key Laboratory of Nuclear Physics and Ion-beam Application (MOE), Institute of Modern Physics, Fudan University, Shanghai 200433, China}
\author{H. Yan}
\email[Contact author: ]{yanhaiyang@nbu.edu.cn}
\affiliation{Institute of Fundamental Physics and Quantum Technology, and School of Physical Science and Technology, Ningbo University, Ningbo, Zhejiang 315211, China}
\maketitle
Muons are probably the most suspicious fermions in which new physics might be involved. The charge radius puzzles of muonic hydrogen~\cite{antognini2013s} and muonic deuterium~\cite{Pohl2016s} are well-known examples.
Extensive experimental efforts have been made to search for exotic spin-dependent interactions. To date, nearly all known searches for such interactions have focused on protons, neutrons, and electrons, while constraints involving other fermions-such as muons-remain scarce~\cite{safronova2018rmp, cong2025a}. Nevertheless, muons stand out as especially suspicious candidates for possible new interactions, given their participation in several outstanding puzzles in modern physics. Parity-violating muonic interactions mediated by new massive gauge bosons in the MeV$\sim$GeV range have even been proposed to explain the proton charge radius puzzle~\cite{Batell2011prl, Barger2012prl}.
The absence of any confirmed new interactions among electrons, neutrons, or protons further suggests that the muon could be a unique and promising probe for long-range new forces. These hypothetical forces might be muonic, meaning they couple exclusively to muons. Notably, such exotic spin-dependent interactions could induce a subtle modification to the muon anomalous magnetic moment by altering the electromagnetic vertex~\cite{yan2019}.

Probing these new interactions is effectively equivalent to searching for axions, which could help resolve several fundamental open questions in modern physics. On one hand, axions are strong candidates for dark matter-one of the most profound unsolved problems in both particle physics and astrophysics~\cite{PDG2024, Fu2017prl}. On the other hand, axions have attracted significant attention in high-energy physics for providing a compelling mechanism to preserve CP symmetry in strong interactions~\cite{Kim2010rmp}.
A variety of new particles predicted by extensions beyond the Standard Model are also potential dark matter candidates, including the axion-like particles (ALPs)~\cite{sikivie2021RMP}, $Z^\prime$ bosons~\cite{Fayet1980aPLB, Fayet1980PLB, Nobuchika2020PLB}, and dark photons~\cite{Holdom1986PLB}.

Recently, significant progress has been made in determining the anomalous magnetic moment $a_\mu = (g-2)/2$ of the muon, providing an important opportunity to probe exotic interactions between muons mediated by new particles.
The anomalous magnetic moments of fundamental particles, such as the electron and muon, are among the most precisely measured quantities in physics.
Any deviation between these experimental results and theoretical predictions may signal the presence of new physics, given the crucial role these quantities have played in the historical development of quantum theory.
The anomalous magnetic moment of the electron has been measured with a precision reaching 1 part in $10^{12}$~\cite{Fan2023PRL}, thereby placing stringent limits on possible exotic interactions coupling to the electron~\cite{yan2019, Andreev1021PRD}.
In contrast, the muon anomaly has been determined with lower precision and, until recently, exhibited a long-standing tension between experimental measurements and theoretical expectations.

Specifically, exotic interactions between muons $\psi$ may be mediated by a new scalar particle $\phi$ or a new vector particle $X$ through the following couplings~\cite{moody1984PRD, dobrescu2006JHP, PDG2024}:
\begin{equation}
\mathcal{L} = \bar{\psi}\phi(g_S + i g_P \gamma_5)\psi + \bar{\psi}X_\mu\gamma^\mu(g_V + g_A \gamma_5)\psi,
\end{equation}
\noindent where $g_i$ (with $i =$ S, P, V, A) denote the dimensionless coupling strengths corresponding to scalar, pseudoscalar, vector, and axial-vector interactions, respectively.
At the one-loop level, the Feynman diagrams derived from this Lagrangian generate corrections to the muon's magnetic dipole moment.
Since the magnetic interaction conserves parity, only the SS, PP, VV, and AA coupling terms contribute.
The total contribution from these four types of couplings to the anomalous magnetic moment is calculated as follows~\cite{Fayet2007PRD, leveille1978NPB, yan2019}:
\begin{widetext}
    \begin{equation}
    \begin{aligned}
        \delta a_\mu&=\frac{1}{8\pi^2}\int_0^1\Big[g_S^2\frac{(1-x)^2(1+x)}{(1-x)^2+x (m_\phi/m_\mu)^2}-g_P^2 \frac{(1-x)^3}{(1-x)^2+x(m_\phi/m_\mu)^2}\Big]dx\\
        &+\frac{m_\mu^2}{4\pi^2m^2_{X}}\int_0^1\Big[g_V^2\frac{x^2(1-x)}{1-x+x^2(m_\mu/m_{X})^2}-g_A^2\frac{x(1-x)(4-x)+2x^3(m_\mu/m_{X})^2}{1-x+x^2(m_\mu/m_{X})^2}\Big]dx,
    \end{aligned}
    \label{deltamu}
\end{equation}
\end{widetext}
where $x$ is the Feynman integration parameter, and $m_\mu$, $m_\phi$, and $m_{X}$ denote the masses of the muon, the new scalar particle, and the new vector particle, respectively.
In this framework, the muon anomaly shows an enhanced sensitivity to new physics compared with the electron, as the contribution to the anomaly scales proportionally with the muon mass~\cite{Fayet2007PRD, Roberts2011JP}.

Using the latest result $\Delta a_\mu = a_\mu^{\rm exp} - a_\mu^{\rm SM} = (38 \pm 63) \times 10^{-11}$~\cite{muoncoll2025, aliberti2025arxiV}, updated constraints on the four couplings can be derived within a Bayesian framework:
\begin{equation}
p(\delta a_\mu|\Delta a_\mu)=\frac{\mathcal{L}(\Delta a_\mu|\delta a_\mu)p(\delta a_\mu)}{\int \mathcal{L}(\Delta a_\mu|\delta a_\mu)p(\delta a_\mu)d\delta a_\mu},
\end{equation}
where $p(\delta a_\mu)$ denotes the prior probability density of the anomaly induced by exotic interactions. It is assumed to be uniformly distributed over the positive interval for the SS and VV couplings, and over the negative interval for the PP and AA couplings, reflecting their opposite signs in contributing to the anomaly~(\ref{deltamu}). The likelihood function $\mathcal{L}(\Delta a_\mu|\delta a_\mu) \sim \mathcal{N}(\mu,\sigma^2)$ follows a Gaussian distribution with $\mu = 38 \times 10^{-11}$ and $\sigma = 63 \times 10^{-11}$, as determined by the experimental measurement of $\Delta a_\mu$.
Consequently, the upper limits on the SS, PP, VV, and AA couplings at the $95\%$ confidence level (CL) can be expressed as follows:
{\small
\begin{equation}
\frac{\int_{0}^{\pm|\delta a_\mu|}\frac{1}{\sqrt{2\pi}\sigma}e^{-\frac{(x-\mu)^2}{2\sigma^2}}dx}{\int_{0}^{\pm\infty}\frac{1}{\sqrt{2\pi}\sigma}e^{-\frac{(x-\mu)^2}{2\sigma^2}}dx}\leq95\%,
\end{equation}}\noindent
where ``$+$'' and ``$-$'' correspond to the positive and negative contribution cases, respectively.
The denominator in the integration serves as a renormalization factor that accounts for the truncation of the probability distribution-specifically, for positive (negative) contribution cases, the negative (positive) side of the distribution is excluded.
In this analysis, we further assume that only one type of contribution is considered at a time, with all others set to zero, as is commonly adopted in similar studies~\cite{baruch2024prl, wei2022nc}.
\begin{figure}[h!]
    \centering
    \subfigure{\includegraphics[width=0.9\linewidth]{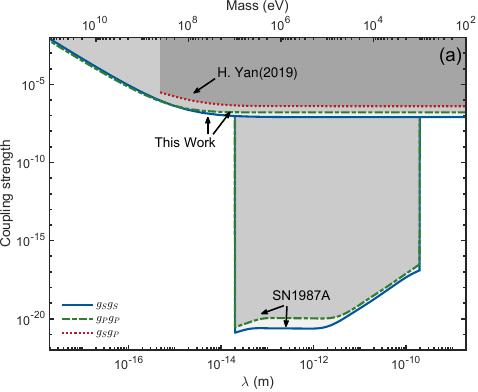}
    }
    \subfigure{\includegraphics[width=0.9\linewidth]{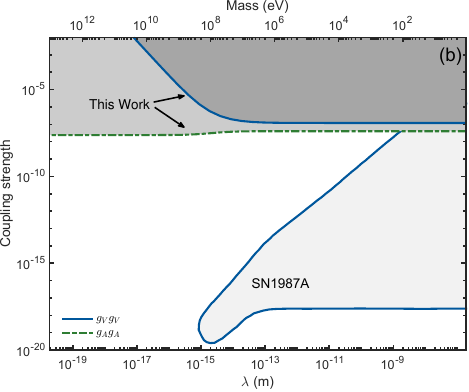}
    }
    \caption{Constraints on exotic spin-dependent interactions involving muons at $95\%$-CL. The gray regions indicate the excluded parameter space. $\lambda = \hbar / (m_{\phi,X} c)$ represents the interaction range, where $\hbar$ is Planck’s constant and $c$ is the speed of light. (a) Constraints on scalar-particle-mediated exotic interactions. The red dotted line represents the result from Ref.~\cite{yan2019}. (b) Constraints on vector-particle-mediated exotic interactions. Constraints on the SS and PP couplings from SN1987A are taken from Ref.~\cite{Caputo2022PRD}, while the VV coupling constraint is from Ref.~\cite{Croon2021JHEP}.}
    \label{fig1}
\end{figure}

The obtained constraints are presented in Fig.~\ref{fig1}, together with those derived from SN1987A observations~\cite{Caputo2022PRD, Croon2021JHEP}.
The blue solid and green dash-dotted curves show the $95\%$-CL limits on $|g_S^\mu|^2$, $|g_P^\mu|^2$, and $|g_A^\mu|^2 m_\mu^2 / m_X^2$ obtained in this work.
The rescaling factor $m_\mu^2 / m_X^2$ in the AA coupling case accounts for the contribution from the longitudinal polarization mode of the vector mediator $X$, which becomes divergent in the limit $m_X \to 0$~\cite{fadeev2022}.
To the best of our knowledge, these results constitute the first laboratory constraints on the SS, PP, VV, and AA couplings involving muons.

For the scalar mediator, the constraints on the SS and PP couplings for interaction ranges $\lambda \gtrsim 10^{-14}$~m are
\begin{equation}
\begin{aligned}
|g_S^\mu|^2 &< 8.0\times 10^{-8}~(95\%{\rm-CL}),\\
|g_P^\mu|^2 &< 1.6\times 10^{-7}~(95\%{\rm-CL}).
\end{aligned}
\end{equation}
The SP coupling, which violates both parity and time-reversal symmetries, can be constrained by the experimental upper limit on the muon electric dipole moment (EDM), as shown by the red dotted line in Fig.~\ref{fig1}(a)~\cite{yan2019}.
For the vector mediator, our constraint on the VV coupling, shown by the blue solid line in Fig.~\ref{fig1}(b), is
\begin{equation}
|g_V^\mu|^2 < 1.2\times 10^{-7}~(95\%{\rm-CL}),
\end{equation}
for $\lambda \gtrsim 10^{-14}$~m.
The behavior of the AA coupling constraint as a function of the mediator mass can be understood as follows.
In the limiting cases of either $m_\mu \ll m_X$ or $m_\mu \gg m_X$, Eq.~(\ref{deltamu}) shows that the AA contribution to the muon anomaly scales as $m_\mu^2 / m_X^2$.
Consequently, the quantity $|g_A^\mu|^2 m_\mu^2 / m_X^2$ becomes independent of the mediator mass $m_X$ in these limits.

In conclusion, we have derived updated constraints on exotic interactions involving muons by utilizing the latest precision measurements of the muon anomalous magnetic moment, $\Delta a_\mu$.
Laboratory-produced muons are naturally polarized and have long been employed in mature experimental techniques such as muon spin rotation ($\mu$SR) spectroscopy, which offers a highly sensitive probe of spin-dependent effects~\cite{yan2019,an2025cpl}.
Adapting and extending $\mu$SR methodologies to search for exotic muon interactions is therefore both experimentally feasible and timely.
Comprehensive exploration of these possibilities across diverse muon-based experimental platforms could fill a major gap in current searches and potentially uncover an entirely new sector of fundamental physics.

\section*{Acknowledgments}
We acknowledge support from the National Natural Science Foundation of China under
grant U2230207.

%

\end{document}